\documentclass[lettersize,journal]{IEEEtran}
\usepackage{amsmath,amssymb,amsfonts}
\usepackage{algorithmic}
\usepackage{algorithm}
\usepackage{array}
\usepackage[caption=false,font=normalsize,labelfont=sf,textfont=sf]{subfig}
\usepackage{textcomp}
\usepackage{stfloats}
\usepackage{url}
\usepackage{verbatim}
\usepackage{graphicx}
\usepackage{cite}
\usepackage{amsmath,amsfonts}
\usepackage{algorithmic}
\usepackage{algorithm}
\usepackage{array}
\usepackage[caption=false,font=normalsize,labelfont=sf,textfont=sf]{subfig}
\usepackage{textcomp}
\usepackage{stfloats}
\usepackage{url}
\usepackage{verbatim}
\usepackage{graphicx}
\usepackage{cite}
\usepackage{adjustbox}
\usepackage{diagbox}
\usepackage[table,dvipsnames]{xcolor}
\DeclareMathAlphabet{\mathcal}{OMS}{cmsy}{m}{n}
\usepackage{scalerel}
\newcommand\mydots{\hbox to 1em{.\hss.\hss.}}

\DeclareMathOperator*{\argmax}{\mathrm{argmax}}
\usepackage{mathtools}
\usepackage{multirow}

\begin{document}

\title{Data-Driven Symbol Detection for Intersymbol Interference Channels with Bursty Impulsive Noise}

\author{Boris Karanov,~\IEEEmembership{Member,~IEEE,}
Chin-Hung Chen,~\IEEEmembership{Student Member,~IEEE,}
Yan Wu, Alex Young, \\ and Wim van Houtum,~\IEEEmembership{Senior Member,~IEEE}
\thanks{Manuscript received xx, xxxx; revised xx, xxxx. (\textit{Corresponding author: Boris Karanov})}
\thanks{B. Karanov, C.-H. Chen and W. v. Houtum are with the Signal Processing Systems (SPS) Group, Electrical Engineering Department, Eindhoven University of Technology, 5600MB Eindhoven, The Netherlands. All authors are also with NXP Semiconductors, High Tech Campus 46, 5656AE Eindhoven, The Netherlands. \textit{(e-mails: b.p.karanov@tue.nl; c.h.chen@tue.nl; yan.wu 2@nxp.com; alex.young@nxp.com; w.j.v.houtum@tue.nl)}}
}

\markboth{Preprint}%
{Karanov \MakeLowercase{\textit{et al.}}: Data-Driven Symbol Detection for Intersymbol Interference Channels with Bursty Impulsive Noise}


\maketitle

\begin{abstract}
We developed machine learning approaches for data-driven trellis-based soft symbol detection in coded transmission over intersymbol interference~(ISI) channels in presence of bursty impulsive noise~(IN), for example encountered in wireless digital broadcasting systems and vehicular communications. This enabled us to obtain optimized detectors based on the Bahl-Cocke-Jelinek-Raviv~(BCJR) algorithm while circumventing the use of full channel state information (CSI) for computing likelihoods and trellis state transition probabilities. First, we extended the application of the neural network (NN)-aided BCJR, recently proposed for ISI channels with additive white Gaussian noise~(AWGN). Our investigation highlighted that, although suitable for estimating likelihoods via labeling of transmission sequences, the BCJR-NN method does not provide a framework for learning the trellis state transitions. In addition to detection over the joint ISI and IN states we also focused on another scenario where trellis transitions are not trivial: detection for the ISI channel with AWGN with inaccurate knowledge of the channel memory at the receiver. Our results showed that, without access to the accurate state transition matrix, the BCJR-NN performance significantly degrades in both settings. To this end, we devised an alternative approach for data-driven BCJR detection based on the unsupervised learning of a hidden Markov model (HMM). The BCJR-HMM allowed us to optimize both the likelihood function and the state transition matrix without labeling. Furthermore, we demonstrated the viability of a hybrid NN and HMM BCJR detection where NN is used for learning the likelihoods, while the state transitions are optimized via HMM. While reducing the required prior channel knowledge, the examined data-driven detectors with learned trellis state transitions achieve bit error rates close to the optimal full CSI-based BCJR, significantly outperforming detection with inaccurate CSI.
\end{abstract}


\section{INTRODUCTION}
\IEEEPARstart{M}{achine} learning (ML) is revolutionizing various aspects in the design of the next-generation communication systems by enabling powerful tools for data-driven optimization. A particularly appealing application area, subject to extensive research in recent years, is the development of advanced ML-based digital signal processing (DSP) algorithms for communication scenarios where the optimal transceiver DSP is not available, for e.g., due to the lack of sufficient channel knowledge and/or implementation complexity constraints~\cite{Osvaldo_ML}. More specifically, the vast majority of recent approaches exploit the universal function approximation capabilities of neural networks (NN) and deep learning (DL)~\cite{Goodfellow,Hornik}. For example, end-to-end DL methods for the joint optimization of the DSP chain via a fully NN-based transceiver has been recently considered and experimentally verified for both wireless~\cite{OShea_Hoydis,Dorner_ten_Brink} and optical fiber communications~\cite{Karanov_1,Karanov_2,Karanov_5}. Nevertheless, continuous end-to-end learning on an actual transmission link requires advanced solutions for optimization of the NN transmitter which introduces substantial complexity overhead~\cite{Faycal,Karanov_3}. The most common application of NNs and DL for signal processing in communication systems is for the data-driven optimization of a specific receiver DSP function, e.g., for equalization, detection or decoding~\cite{Hager,Farsad,Nachmani}. Training of an NN at the receiver side does not require the backpropagation of gradients through the physical channel and is thus a relatively less complex task which typically involves supervised learning via labeled transmission sequences. For instance, the authors of~\cite{Farsad} proposed an advanced bidirectional recurrent NN~(BRNN) sequence detection scheme which performs close to the optimal maximum likelihood sequence detector. Notably, to alleviate the computational demands of employing a fully neural network-based DSP module, model-based deep learning approaches have been considered, where a simple NN (e.g. a feedforward NN) augments a conventional signal processing algorithm. For example an NN-aided digital fiber nonlinearity compensation scheme was proposed in~\cite{Hager}. More recently, for the task of symbol detection, NN-based approaches for data-driven implementation of the trellis-based Viterbi~\cite{Forney} and Bahl-Cocke-Jelinek-Raviv~(BCJR)~\cite{BCJR_74} algorithms were proposed~\cite{Nir_model_based,Nir_BCJRNet,Nir_ViterbiNet}. Investigated for finite-state stationary intersymbol interference (ISI) channels with additive white Gaussian noise~(AWGN), these detectors achieved error rate performances close to the optimal full channel state information~(CSI)-based BCJR and Viterbi processing. Importantly, these receivers enabled learning of channel likelihoods directly from labeled transmission sequences, circumventing the often impractical requirement for perfect CSI. Moreover, the authors showed that the NN-based BCJR and Viterbi symbol detectors present a superior data-driven alternative in terms of training data utilization and computational complexity compared to the state-of-the-art BRNN receiver.

In more detail, the optimal symbol detector based on the maximum a posteriori probability (MAP) criterion can be realized via BCJR --- an efficient message-passing algorithm which uses channel likelihoods and state transition probabilities to compute posterior beliefs. As proposed in~\cite{Nir_BCJRNet}, the NN-based BCJR (denoted BCJR-NN in our paper) combines a small feedforward NN, trained via received sequences labeled with the transmitted symbols, and a Gaussian mixture density estimator to compute the channel likelihoods for the BCJR algorithm. The scheme requires state transition probabilities to be provided as additional domain knowledge or separately counted in the labeled data. 

However, there are communication channel models where transitions are not trivial and optimizing them in a data-driven manner is particularly suitable. A practically-relevant example is symbol detection over a finite-state ISI channel with AWGN where the symbol memory of the BCJR algorithm is greater than the channel memory --- a scenario that occurs in real-world systems with over-estimated states for ``worst-case'' channel conditions. Another example is the transmission over ISI channels affected by impulsive noise~(IN). Such noise can be found in powerline communications, digital broadcasting systems as well as vehicular communications~\cite{Shan, Sanchez_TV,Mostafa,Yang,Oh,Liu}. A commonly used assumption in such systems is the Middleton Class A noise model~\cite{Middleton,Middleton_new}. More recently, to account for the bursty behavior of the IN, the Markov-Middleton noise model has been proposed~\cite{Ndo_MiddletonMarkov}. Such a finite-state model exhibits a non-trivial state transition matrix governed by impulsive noise parameters. The performance of message-passing algorithms such as BCJR have been investigated for optimal detection over the Markov-Middleton channel, showing significant improvement over detection based on the AWGN assumption~\cite{Ndo_MiddletonMarkov,Mirbadin_Middleton_2}. Note that in these cases full prior knowledge of the IN parameters is required for obtaining the optimal detector.

Our work focused on the machine learning perspective of BCJR symbol detection 1) when knowledge of the channel memory is inaccurate, and 2) over joint ISI and impulsive noise states. Developing data-driven solutions for these communication scenarios is important because they yield a non-trivial state transition matrix for the trellis-based detector which is often unavailable or impractical to provide as prior knowledge. Moreover, labeling of the channel states is also difficult, especially in the presence of IN whose source is typically external to the system. This prompted us, in addition to extending the application of the NN-based BCJR, to consider a fundamentally different perspective by interpreting the learnable detector as a hidden Markov model (HMM)~\cite{Rabiner_HMM}. In particular, we optimized the observation likelihood function and the state transition matrix parameters of the HMM in an unsupervised fashion using received symbol sequences and the Baum-Welch algorithm~\cite{Baum_72,Welch_BaumWelch}. These were then utilized directly for BCJR symbol detection. Expectation-maximization approaches such as the Baum-Welch algorithm have previously been used for learning likelihoods in ISI communication channels with AWGN for example in~\cite{Kaleh} and more recently in~\cite{Schmalen} where the authors investigated joint channel estimation and symbol detection schemes. Conversely, designing data-driven detection for the applications of interest in our study puts a strong emphasis on learning the trellis state transition probabilities. Furthermore, we investigated the feasibility of a hybrid HMM-NN detection approach to BCJR (BCJR-HMM-NN) where the NN-computed likelihoods are used together with an HMM-optimized state transition matrix. 

This paper is organized as follows: Sec.~\ref{sec:hmm} gives a comprehensive overview of HMMs --- a perspective we adopt for interpreting the proposed data-driven BCJR detector designs. Next, the communication channel model is presented in Sec.~\ref{sec:channel}. The proposed machine learning-based detection schemes are described in Sec.~\ref{sec:system}, while their numerical performance is reported in Sec.~\ref{sec:results}. Section~\ref{sec:concl} concludes the paper.

\section{HMMs and Computing Posterior Beliefs via the Forward-Backward Algorithm}\label{sec:hmm}
\begin{figure}[t!]
	\centering
	\includegraphics[width=0.95\columnwidth, keepaspectratio=true]{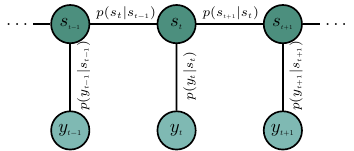}
	\caption{\label{fig:HMM_schematic_1} HMM representation as an undirected graph showing the factorization and conditional independencies implied by the model.}
\end{figure}
An HMM is a statistical model that captures the temporal behavior of a sequence of latent variables (\emph{states}) from a Markov process via a sequence of state-dependent observable variables (\emph{observations})~\cite{Rabiner_HMM,Jurafsky_ASR}. To highlight the conditional independencies and factorization implied by an HMM, Fig.~\ref{fig:HMM_schematic_1} shows its representation as an undirected graph. The model consists of an observed sequence $\left(\ldots,y_{t-1},y_t,y_{t+1}\ldots\right)$ where at a given time $t$ an observation $y_t$ is produced by a state $s_t$. In particular, for the state and observation sequences of length $T$, the HMM represents the factorization
\begin{equation*}
    \begin{split}
	   p(s_1,s_2,\mydots,s_T,y_{1},y_2,\mydots,y_T)= &p(s_1)p(y_1|s_1) \\ & \times \prod_{t=2}^{T}p(s_t|s_{t-1})p(y_t|s_t).
    \end{split}
\end{equation*}
We see that the probability of an observation at a given time depends only on the state that produced it, such that
\begin{equation*}
	p(y_t|s_1,\mydots,s_{t},\mydots,s_T,y_{1},\mydots,y_{t},\mydots,y_T) = p(y_t|s_t).
\end{equation*}
Furthermore, the Markovianity of the underlying state generation process dictates that the probability of being at state $s_t$ at time $t$ depends only on the state at the previous time:
\begin{equation*}
	p(s_t|s_1,\mydots,s_{t-1}) = p(s_t|s_{t-1}).
\end{equation*}

\begin{figure}[t!]
	\centering
	\includegraphics[width=\columnwidth, keepaspectratio=true]{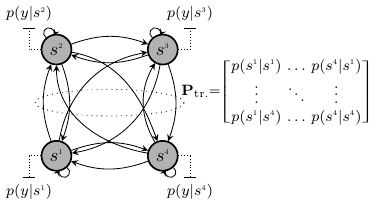}
	\caption{\label{fig:HMM_schematic_2} State diagram representation of a 4-state HMM highlighting the model parameters: the observation likelihood function and the state transition matrix.}
\end{figure}

It is worth mentioning here that, for the purposes of the presented work, the HMM is specified over a finite set of $Q$ possible discrete states $\mathcal{S}=\{s^1,\ldots,s^Q\}$, while the observation variables follow a continuous distribution. Figure~\ref{fig:HMM_schematic_2} shows a state diagram representation of an HMM. The model consists of a \emph{state transition matrix} parameter
\begin{equation*}
	\mathbf{P}_{\text{tr.}}=\begin{bmatrix}p(s^{1}|s^{1})&\hdots&p(s^{Q}|s^{1})\\ \vdots & \ddots & \vdots\\ p(s^{1}|s^{Q})&\hdots&p(s^{Q}|s^{Q})\end{bmatrix},
\end{equation*}
where each element represents the probability $p(s^j|s^i)$ of moving from state $s^i$ to $s^j$ in one time step. The second HMM parameter is the observation likelihood function $p(y|s^j)$ describing the probability of an observation being generated from a state $s^j$. Note that the characterization of an HMM also includes a distribution $p(s)$ over the initial states which can be obtained from the transmission model or learned together with the other HMM parameters.

\begin{figure}[t!]
	\centering
	\includegraphics[width=0.95\columnwidth, keepaspectratio=true]{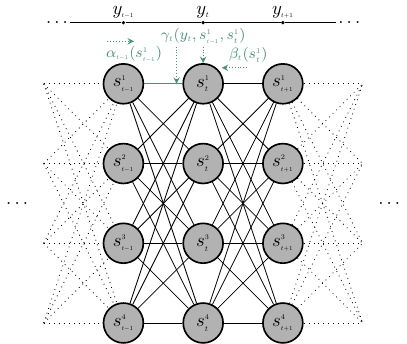}
	\caption{\label{fig:HMM_schematic_3} Trellis diagram representation of a 4-state HMM illustrating the message passing mechanism in the forward-backward algorithm.}
\end{figure}

The HMM is a tree-structured graph and the sum-product message-passing algorithm (belief propagation) can be used to perform efficient inference~\cite{Bishop_Pattern}. In particular, the belief propagation variant specialized to the structure of an HMM is called the \emph{forward-backward algorithm}~\cite{Rabiner_HMM,Jurafsky_ASR}. In the following, we describe the procedure for obtaining the marginal distribution for a latent state given the full sequence of observations, i.e., computing the posterior belief
\begin{equation}\label{eq:post_belief_state}
\begin{split}
	p(s_t|y_1,y_2,\mydots, y_T)& \propto p(y_1,y_2,\mydots, y_T,s_t) \\ & =\sum_{s_{t-1}}p(y_1,y_2,\mydots, y_T,s_{t-1},s_{t}).
\end{split}
\end{equation}
Using the chain rule of probability and the conditional independencies in the HMM graphical model, we can write the joint distribution from the right-most side of~\eqref{eq:post_belief_state} as
\begin{equation}\label{eq:HMM_factoriz_tr}
	\begin{split}
		p(y_1,y_2,\mydots, y_T,s_{t-1},s_{t})= & p(y_1,y_2,\mydots,y_{t-1},s_{t-1}) \\ & \times p(s_{t}|s_{t-1})p(y_t|s_t)\\ & \times p(y_{t+1},y_{t+2},\mydots,y_T|s_{t}).
	\end{split}
	\raisetag{1\normalbaselineskip}
\end{equation}
We first identify a \emph{branch metric} term defined as
\begin{equation}\label{eq:gamma_branch}
	\gamma_t(y_{t},s_{t-1},s_{t}) = p(s_{t}|s_{t-1})p(y_t|s_t).
\end{equation}
Moreover,~\eqref{eq:HMM_factoriz_tr} entails a \emph{forward} probability which, exploiting the factorization associated with the graph structure, can be recursively updated via messages $\alpha$ defined as
\begin{equation}\label{eq:alpha_msg}
	\begin{split}
		\alpha_{t}(s_t)&=p(y_1,y_2,\mydots,y_{t},s_t)  \\ & =\sum_{s_{t-1}}\alpha_{t-1}(s_{t-1})\gamma_t(y_{t},s_{t-1},s_{t}),
	\end{split}
\end{equation}
for $\forall t\!=\!\{2,\mydots,T\}$, where $\alpha_1(s_1)=p(y_1,s_1)$. Correspondingly, \emph{backward} probability is computed via messages $\beta$ as
\begin{equation}\label{eq:beta_msg}
	\begin{split}
		\beta_{t}(s_{t})&=p(y_{t+1},y_{t+2}\mydots,y_T|s_{t}) \\  
		&=\sum_{s_{t+1}}\gamma_t(y_{t+1},s_{t},s_{t+1}) \beta_{t+1}(s_{t+1}),
	\end{split}
\end{equation}
for $\forall t=\{1,\mydots,T-1\}$ with $\beta_T(s_T)=1$. 
Substituting~\eqref{eq:gamma_branch},~\eqref{eq:alpha_msg} and~\eqref{eq:beta_msg} in~\eqref{eq:HMM_factoriz_tr}, we can now write the joint distribution of interest as the product of the forward and backward probabilities and the branch metric:
\begin{equation}\label{eq:forw_backw_BCJR}
	p(y_1,y_2,\mydots, y_T,s_{t-1},s_{t})\!=\!\alpha_{t-1}(s_{t-1})\gamma_t(y_t,s_{t-1},s_{t})\beta_{t}(s_{t}).
\end{equation}
To illustrate this message-passing process, Fig.~\ref{fig:HMM_schematic_3} shows a trellis diagram representation of an HMM. From a communication system perspective this ``$\alpha$-$\beta$-$\gamma$'' forward-backward method for computation of the posterior beliefs results in the well-known \emph{BCJR algorithm}~\cite{BCJR_74} which we used for maximum a-posteriori~(MAP) symbol detection as further explained in Sec.~\ref{sec:system}.
\section{Communication Channel}\label{sec:channel}
In this paper we consider a wireless communication channel subject to impulsive interference which can be found for example in digital broadcasting systems and vehicular communications~\cite{Sanchez_TV,Yang,Oh,Liu}. The channel we consider is characterized by the presence of inter-symbol interference~(ISI) due to the multipath signal propagation and bursty impulsive noise whose power fluctuates in time. In particular, the input-to-output relationship for the channel is given by
\begin{equation}\label{eq:channel_io}
	y_{t} = \sum_{l=1}^L h_{l} x_{t-l+1} + z_t,
\end{equation}
where $x_t\in\mathcal{X}$ is the transmitted symbol at time $t\in\{1,2,\ldots T\}$ from a constellation set $\mathcal{X}$, with $T$ being the transmission frame length. The channel output at time $t$ is denoted by $y_t$. The ISI response is given by the set of taps $\{h_{1},...,h_{L}\}$ with $L$ being the finite ISI memory. We assume an additive interference component which is independent of the transmitted symbols and denoted by $z_t$.

\subsection{Intersymbol Interference}
We adopt the ISI model from \cite{Nir_BCJRNet,Nir_model_based,Chen_1} where the application of data-driven NN-based BCJR symbol detection has been investigated. In this model the channel coefficients follow an exponentially decaying profile and, as in \cite{Chen_1}, are normalized to unit power, i.e.,
\begin{equation*}
	h_{l} = \frac{e^{-\eta(l-1)}}{\sqrt{\sum_{l=1}^{L}|e^{-\eta(l-1)}|^2}},\qquad l={1,...,L},
\end{equation*}
where $\eta$ determines the decay rate. From~\eqref{eq:channel_io} we can identify $|\mathcal{X}|^L$ possible ISI states $\mathbf{s}^{\dagger}_t\triangleq\left(x_{t},x_{t-1},\mydots,x_{t-L+1}\right)$. ISI state transitions are driven by the transmitted symbols, which are assumed equiprobable, i.e., symbol $x_t$ is transmitted at time $t$ with probability $\frac{1}{|\mathcal{X}|}$. Therefore, the probability of being at state $\mathbf{s}^{\dagger}_t$ at time $t$ depends only on the previous state $\mathbf{s}^{\dagger}_{t-1}$ as
\begin{equation} \label{eq:ISI_tr}
	\begin{split}
		&p(\mathbf{s}^{\dagger}_t|\mathbf{s}^{\dagger}_{t-1})=\begin{cases}
			\frac{1}{|\mathcal{X}|}, &(\mathbf{s}^{\dagger}_t, \mathbf{s}^{\dagger}_{t-1}) \in \mathcal{S}^{\dagger}\\
			0, & \text{otherwise}.
		\end{cases}
	\end{split}
\end{equation}
Here $\mathcal{S}^{\dagger}$ denotes the pairs $(\mathbf{s}^{\dagger}_t,\mathbf{s}^{\dagger}_{t-1})$ corresponding to all state transitions driven by the transmitted symbol $x_t$. Thus, in the absence of other effects, the ISI-affected symbol for each state becomes
\begin{equation}\label{eq:smean}
	\mu_{\mathbf{s}^{\dagger}_t} = \sum_{l=1}^L h_{l} x_{t-l+1},
\end{equation}

This is a finite-state time-invariant model of the ISI. In~\cite{Nir_BCJRNet,Nir_model_based}, the authors extended this model to account for channel taps with random variations in time. This was done by introducing a Gaussian-distributed time-varying distortion to $h_l$ with zero mean and variance $\sigma_h^2$ in presence of which the ISI-affected symbol can be expressed as
\begin{equation}\label{eq:smean_}
	\tilde{\mu}_{\mathbf{s}^{\dagger}_t} = \sum_{l=1}^L \left(h_{l} + \epsilon_{l,t}\right) x_{t-l+1},
\end{equation}
where $\epsilon_{l,t} \sim \mathcal{N}(0,\sigma_{h}^2)$.
In~\cite{Nir_BCJRNet} such a model for the received symbols was assumed during the training stage of the NN-based BCJR detector as well as for likelihood computations in the conventional BCJR detection and it was referred to as \emph{CSI uncertainty}. In~\cite{Chen_1} further investigation on the performance of the BCJR detection scheme for rapidly time-varying ISI channels with AWGN was carried out. In our work we also considered such an ISI scenario in the presence of impulsive interference.
\subsection{Impulsive Interference}
We model the behavior of the interference $z_t$ using the Markov-Middleton impulsive noise model~\cite{Ndo_MiddletonMarkov} which allows us to account for the bursty (correlated) nature of the impulsive noise events. Recently, the impulsive behavior of electric vehicle (EV)-induced interference, for e.g., originating from an EV power converter or a switch, in the digital audio broadcasting (DAB) band was investigated on a real world test-bed, showing a good agreement with the Markov-Middleton noise behavior~\cite{Chen_2}. In particular, under the Markov-Middleton model the zero-mean noise sample $z_t$ is distributed as
\begin{equation*}
	p(z_t) = \sum_{j=0}^{N-1}\frac{p_j} {\sqrt{2\pi \sigma^2_{j}}}\exp\left(-\frac{z_{t}^2}{2\mathbf{\sigma}^2_{j}}\right),
\end{equation*}
where
\begin{equation}\label{eq:middl_p_var}
	p_j = \frac{e^{-A}A^{j}/j!} {\sum_{j=0}^{N-1} e^{-A}A^{j}/j!},\hspace{0.2cm} \text{and} \hspace{0.2cm} \sigma^2_j = \sigma^2\frac{j/A+\Gamma}{1+\Gamma},
\end{equation}
with $N$ denoting the number of noise levels. Here we associate with an interference state $\tilde{s}^{j}$ the variance $\sigma^2_{\tilde{s}^j}$ for the $j$-th noise level given by the right-hand side of~\eqref{eq:middl_p_var}. The model parameters $A$, $\Gamma$, and $\sigma^2$ represent the impulsive index, the background-to-impulsive noise ratio and the total power of the interference $z_k$, respectively. Note that since we considered the BPSK modulation format with constellation points $\mathcal{X}=\{-1,1\}$ and normalized the ISI channel taps to unit power, the signal-to-interference-and-noise ratio at the receiver becomes $\text{SINR}=1/\sigma^2$. Also note that $\tilde{s}^{0}$ corresponds to background AWGN. When $N=1$ there is only background noise present in the system and thus we use the signal-to-noise ratio~(SNR) metric. Furthermore, the Markov-Middleton impulsive model involves a parameter $r\in[0,1]$ that establishes correlation between noise samples. In particular, the probability of being in interference state $\tilde{s}_t^{j}$ at time $t$ depends only on the previous state $\tilde{s}_{t-1}^{i}$ with transitions governed by
\begin{equation} \label{eq:IN_tr}
	\begin{split}
		&p(\tilde{s}_t^{j}|\tilde{s}_{t-1}^{i})=\begin{cases}
			r + (1-r)p_j,& \tilde{s}^i=\tilde{s}^j \\
			(1-r)p_j, & \text{otherwise}.
		\end{cases}
	\end{split}
\end{equation}

\section{Receiver Design}\label{sec:system}

In this section we describe the proposed methods for data-driven maximum a posteriori probability (MAP) symbol detection for the ISI channel with bursty IN. The MAP symbol detector is based on the BCJR algorithm and we investigate both NN and HMM approaches for its machine learning-based implementation. We first start with the description of the BCJR algorithm for optimal inference over the joint ISI and IN channel states. Using our notation from the definition of the channel in~Sec.~\ref{sec:channel} we identify a communication system state as the pair of the ISI process state and the impulsive noise process state at a given time instance $t$, i.e, $\mathbf{s}_t\triangleq\left(\mathbf{s}^{\dagger}_t,\tilde{s}_t\right)$. Note that since the ISI and impulse noise processes are assumed independent we obtain a set $\mathcal{S}$ of system states which has the cardinality of $|\mathcal{S}|=Q=N\cdot|\mathcal{X}|^L$, where $N$ is the number of noise levels, $|\mathcal{X}|$ is the cardinality of the set of modulation symbols and $L$ is the ISI symbol memory. Based on the channel model from~\eqref{eq:channel_io}, we can factorize the joint distribution of the sequences of system states and received symbols as
\begin{equation}\label{eq:channel_fact}
	\begin{split}
        p(\mathbf{s}_1,\mathbf{s}_2,\mydots,\mathbf{s}_T,y_{1},y_2,&\mydots,y_T)=p(\mathbf{s}_1)|p(y_1|\mathbf{s}_1) \\ & 	\times\prod_{t=2}^{T}p(\mathbf{s}_t|\mathbf{s}_{t-1})p(y_t|\mathbf{s}_t),
	\end{split}
	\raisetag{1\normalbaselineskip}
\end{equation}
where for the ease of notation we assumed that the received symbol at $t=1$ has a full ISI state. This can be ensured via a guard band of $L-1$ transmitted symbols at the beginning of the transmission frame. Going further, we can express the likelihood of an output symbol given a channel state by
\begin{equation}\label{eq:ch_output_likelihood}
	p(y_t|\mathbf{s}_t) = \frac{1} {\sqrt{2\pi \sigma^2_{\mathbf{s}_t}}}\exp\left(-\frac{\left(y_{t}-\mu_{\mathbf{s}_t}\right)^2}{2\sigma^2_{\mathbf{s}_t}}\right).
\end{equation}
We note that for the considered channel model the received symbol follows a Gaussian distribution $y_t\sim\mathcal{N}(\mu_{\mathbf{s}_t},\sigma^2_{\mathbf{s}_t})$, whose mean and variance parameters are determined by the ISI and the impulsive interference states, i.e.,~\eqref{eq:smean} and~\eqref{eq:middl_p_var}, respectively.
\begin{figure*}[t!]
	\centering
	\includegraphics[width=\textwidth, keepaspectratio=true]{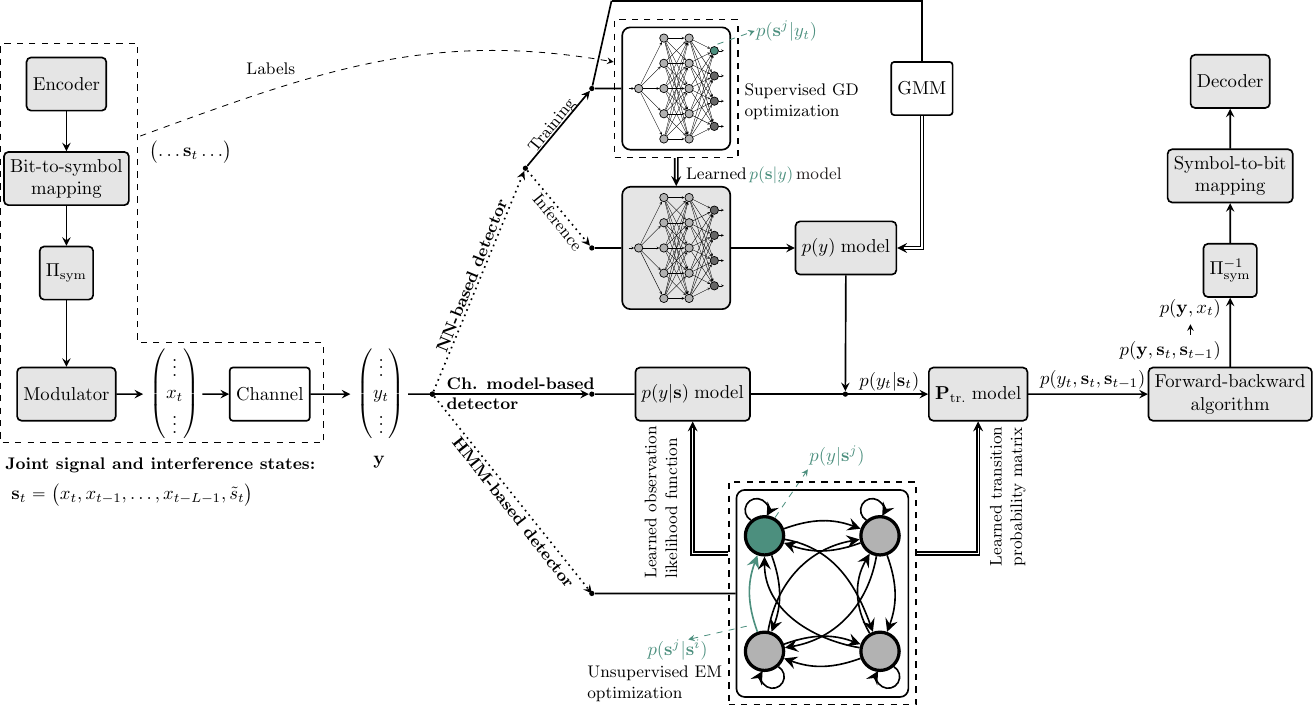}
	\caption{\label{fig:System_schematic} Schematic of the communication system using channel model-based or data-driven symbol detector based on the forward-backward algorithm (BCJR detection). The neural network can be used in the data-driven estimation of the channel likelihoods, while both the channel likelihood estimation and the state transitions can be obtained by an HMM optimization.}
\end{figure*}
We construct a state transition matrix for the system by taking the Kronecker product of the transition matrices for the ISI and the implusive interference defined by~\eqref{eq:ISI_tr} and~\eqref{eq:IN_tr}, respectively. Thus, we obtain the probability of the system being at the joint ISI and IN state $\mathbf{s}_t$ at time $t$ to be governed by the previous system state as
\begin{equation} \label{eq:ISI_IN_tr}
	\begin{split}
		&p(\mathbf{s}_t|\mathbf{s}_{t-1})=\begin{cases}
			\frac{r + (1-r)p_j}{|\mathcal{X}|}, & \hspace{-0.175cm}(\mathbf{s}^{\dagger}_t, \mathbf{s}^{\dagger}_{t-1}) \!\in\! \mathcal{S}^{\dagger} \hspace{0.075cm} \text{and} \hspace{0.075cm} \tilde{s}^i\!=\!\tilde{s}^j\\
			\frac{(1-r)p_j}{|\mathcal{X}|}, & \hspace{-0.175cm}(\mathbf{s}^{\dagger}_t, \mathbf{s}^{\dagger}_{t-1}) \!\in\! \mathcal{S}^{\dagger} \hspace{0.075cm} \text{and} \hspace{0.075cm} \tilde{s}^i\!\neq\!\tilde{s}^j \\
			0, &\hspace{-0.175cm} \text{otherwise}.
		\end{cases}
		\raisetag{1\normalbaselineskip}
	\end{split}
\end{equation}
\subsection{Optimal symbol detection with the BCJR algorithm}
The optimal symbol detector based on the MAP criterion that minimizes the symbol error rate is given by
\begin{equation*}
	\hat{x}_t \!=\! \argmax_{x_t\in \mathcal{X}}p(x_t|y_1,\mydots, y_T) \!=\! \argmax_{x_t\in \mathcal{X}}p(y_1,\mydots, y_T,x_t),
	\raisetag{1\normalbaselineskip}
\end{equation*}
where the joint distribution on the right-hand side of the above equation can be written as
\begin{equation}\label{eq:joint_symb_rec_seq}
	p(y_1,\mydots, y_T,x_t) = \sum_{\mathclap{\mathbf{s}_t, \mathbf{s}_{t-1}\in \mathcal{S}^{\ast}}} p(y_1,\mydots, y_T,\mathbf{s}_{t-1}, \mathbf{s}_{t}).
\end{equation}
For ease of notation we denoted with $\mathcal{S}^{\ast}$ the subset of allowed system state transitions driven by the transmitted symbol $x_t$. Utilizing the factorization from~\eqref{eq:channel_fact}, the right-hand side of~\eqref{eq:joint_symb_rec_seq} can be recursively computed via the $\alpha$-$\beta$-$\gamma$ forward-backward (BCJR) algorithm from~\eqref{eq:forw_backw_BCJR}, which was extensively described in Sec.~\ref{sec:hmm}. Note that the branch metrics as well as the forward and backward probabilities are functions of the joint ISI and IN channel states. Their computation conventionally requires full channel knowledge to compute the state transition probability $p(\mathbf{s}_{t}|\mathbf{s}_{t-1})$ from~\eqref{eq:ISI_IN_tr} as well as the channel likelihood $p(y_t|\mathbf{s}_{t})$ from~\eqref{eq:ch_output_likelihood}. In the following we describe both the channel model-based and the machine learning approaches to performing these computations.

Figure~\ref{fig:System_schematic} shows a schematic of the considered communication system. We investigated the bit error rate performance for coded transmission. At the transmitter the information bits are first encoded using a forward error correcting (FEC) code. The encoded bits are mapped to integer symbols, interleaved and modulated on transmission symbols $x_t$. The symbol sequence is transmitted through the ISI channel with bursty IN whose model we described in Sec.~\ref{sec:channel}. In the presence of full CSI at the receiver, the likelihoods of the noisy symbols in the received sequence can be evaluated directly via~\eqref{eq:ch_output_likelihood}. To obtain the branch metrics, these are then multiplied by the state transition probabilities computed from~\eqref{eq:ISI_IN_tr}. When full CSI is not available, we resort to data-driven BCJR detection.
\subsection{Data-Driven NN-based BCJR Detection}\label{sec:BCJR-NN}
The BCJR-NN detector utilizes the received signal in two ways: In training mode the received noisy symbols $\left(\ldots,y_{t-1},y_t,y_{t+1}\ldots\right)$ from a training sequence are labeled with the corresponding channel states $\mathbf{s}_t=\begin{pmatrix} x_t,x_{t-1},\hdots,x_{t-L-1},\tilde{s}_t\end{pmatrix}$. The training dataset of $T$ examples, formed as $\{y_t,\mathbf{s}_{t}\}_{t=1}^{T}$ is used for supervised optimization of an NN-based estimator for the likelihood $p(y_t|\mathbf{s}_{t})$. In inference mode the learned estimator is used to evaluate the likelihood of the received symbols in transmission sequences generated independently from the training ones. In more detail, the NN is employed to classify channel states using a received symbol, thus yielding an approximation $p^{\theta}(\mathbf{s}_{t}|y_t)$ where $\theta$ denotes the NN parameter set of weight matrices and bias vectors for each layer. Employing a softmax output layer, the parameters $\theta$ are optimized by minimizing the average cross-entropy loss over the training dataset, i.e., 
\begin{equation*}
	\mathcal{L}(\theta) = -\frac{1}{T}\sum_{t=1}^{T}\log{p^{\theta}(\mathbf{s}_{t}|y_t)},
\end{equation*}
via stochastic gradient descent (GD).
 Furthermore, as proposed in~\cite{Nir_BCJRNet}, a Gaussian mixture model is utilized to approximate the marginal distribution $p^\phi(y_t)$ from the received training sequence. The channel likelihoods required for the BCJR algorithm can then be computed as
\begin{equation*}
	p^{\theta,\phi}(y_t|\mathbf{s}_{t}) = \frac{p^{\theta}(\mathbf{s}_{t}|y_t) p^{\phi}(y_t)}{p(\mathbf{s}_{t})}.
\end{equation*}
Note that the marginal probability of the joint channel states should be provided, based on the transmission model. Importantly, in order to carry out the BCJR algorithm, the state transition probabilities $p(\mathbf{s}_{t}|\mathbf{s}_{t-1})$ also need to be available as prior knowledge. Subsequently, the branch metrics for the BCJR detection can be computed. 

\subsection{Data-Driven HMM-based BCJR Detection}\label{sec:BCJR-HMM}
The BCJR-HMM detector utilizes a sequence of unlabeled received symbols $\left(\ldots,y_{t-1},y_t,y_{t+1}\ldots\right)$ to learn the distribution parameters of the likelihood function as well as to optimize the state transition matrix. This machine learning approach views the design of the symbol detector from the prism of the hidden Markov model framework described in Sec.~\ref{sec:hmm}. Given an observation sequence $\left(\ldots,y_{t-1},y_t,y_{t+1}\ldots\right)$ and the number of possible states, we can carry out an optimization for the parameters of an HMM. In such a way we obtain a \emph{data-driven model} of the underlying Markov process of interest. The HMM learning procedure involves a combination of the previously discussed computation of posterior beliefs via the forward-backward algorithm (expectation step) and a maximization step which updates the model parameters. This expectation-maximization method for updating the HMM parameters is also known as the \emph{Baum-Welch algorithm}~\cite{Baum_72}. We start by describing the learning of the state transition probabilities. First, we define the posterior probability of being in state $\mathbf{s}^i$ at time $t-1$ and transitioning to state $\mathbf{s}^j$ at time $t$ given the observed sequence as
\begin{equation*}
	\begin{split}
		\xi_t(\mathbf{s}_{t-1}^i,\mathbf{s}_{t}^j) & = p(\mathbf{s}^i_{t-1},\mathbf{s}^j_{t}|y_1,\mydots, y_T) \\ & =\frac{p(y_1,\mydots, y_T,\mathbf{s}^i_{t-1},\mathbf{s}^j_{t})} {p(y_1,\mydots, y_T)} \\ &=\frac{\alpha_{t-1}(\mathbf{s}^i_{t-1})\gamma_t(y_t,\mathbf{s}^i_{t-1},\mathbf{s}^j_{t})\beta_{t}(\mathbf{s}^j_{t})}{\sum_{j=1}^{Q}\alpha_T(\mathbf{s}^j_T)}.
	\end{split}
\end{equation*}
Note that for the numerator in the final expression we used the factorization of the joint distribution from~\eqref{eq:forw_backw_BCJR}, while in the denominator we employed the forward probability from~\eqref{eq:alpha_msg} computed at time $t=T$ to obtain the marginal probability of the full observation sequence. The estimate of the state transition probability can then be obtained by the ratio of the expected number of transitions from $\mathbf{s}^i$ to $\mathbf{s}^j$ (over all $t$) and the total expected number of transitions from state $\mathbf{s}^i$ as
\begin{equation*}
	\begin{split}
		\hat p(\mathbf{s}^{j}|\mathbf{s}^i) & =\frac{\sum_{t=2}^{T}\xi_t(\mathbf{s}_{t-1}^i,\mathbf{s}_{t}^j)} {\sum_{t=2}^{T}\sum_{j=1}^{Q}\xi_t(\mathbf{s}_{t-1}^i,\mathbf{s}_{t}^j)}.
	\end{split}
\end{equation*}
We now explain the update of the observation likelihood function. For this we are first interested in the posterior belief of being in state $\mathbf{s}^j$ at time $t$ given by
\begin{equation}\label{eq:HMM_BaumWelch_xi}
	\begin{split}
		\xi_t(\mathbf{s}_t^j) & = p(\mathbf{s}^j_t|y_1,\mydots, y_T) \\ & =\frac{\sum_{i=1}^{Q}p(y_1,\mydots, y_T,\mathbf{s}^i_{t-1},\mathbf{s}^j_{t})} {p(y_1,\mydots, y_T)} \\ &=\frac{\sum_{i=1}^{Q}\alpha_{t-1}(\mathbf{s}^i_{t-1})\gamma(y_t,\mathbf{s}^{i}_{t-1},\mathbf{s}^{j}_t)\beta_{t}(\mathbf{s}^j_{t})}{\sum_{j=1}^{Q}\alpha_T(\mathbf{s}^j_T)}.
	\end{split}
\end{equation}
In this work, due to the channel model described in Sec.~\ref{sec:channel}, we considered observation likelihood function that follows a Gaussian distribution. It is worth noting that different choice of distribution (e.g., a Gaussian mixture~\cite{Jurafsky_ASR}) can be made and its parameters can be updated in a similar fashion. Consequently, the likelihood of an observation is given by~\eqref{eq:ch_output_likelihood}. Using~\eqref{eq:HMM_BaumWelch_xi}, the mean of the observation likelihood function for state $\mathbf{s}^j$ can be updated as 
\begin{equation*}
	\begin{split}
		\hat{\mu}_{\mathbf{s}^{j}} = \frac{\sum_{t=1}^{T}\xi_t(\mathbf{s}^j_t)y_t} {\sum_{t=1}^{T}\xi_t(\mathbf{s}^j_t)}.
	\end{split}
\end{equation*}
Similarly, estimation of the variance is performed as
\begin{equation*}
	\begin{split}
		\hat{\sigma}^2_{\mathbf{s}^j} = \frac{\sum_{t=1}^{T}\xi_t(\mathbf{s}^j_t)(y_t-\hat{\mu}_{\mathbf{s}^j})^2} {\sum_{t=1}^{T}\xi_t(\mathbf{s}^j_t)}.
	\end{split}
\end{equation*}

After the symbol detection phase, the soft outputs, expressed by the joint probability from~\eqref{eq:joint_symb_rec_seq}, for the channel model-based as well as the data-driven BCJR detectors are appropriately deinterleaved and fed to the FEC decoder which recovers the information bits. The obtained BER performance of the investigated systems is reported next.

\section{Performance Results}\label{sec:results}

\begin{table}[tpb]
	\caption{Simulation parameters}
	\renewcommand{\arraystretch}{1.3}
	\centering
	\begin{tabular}{|c|c|} 
		\hline
		{FEC scheme} & {$(171,133)$ conv. code} \\
		\hline
		{Code rate} & {$1/2$} \\
		\hline
		{Code memory} & {$7$ bits} \\
		\hline
		{Modulation scheme} & {BPSK ($|\mathcal{X}|=2$)} \\
		\hline
		{Symbol sequence length ($T$)} & {500000} \\
		\hline
		{Monte Carlo iterations} & {500} \\
		\hline
		{Channel memory ($L$)}& $1$ or $2$ sym. \\
		\hline
		{Channel taps decay rate ($\eta$)} & $1$ \\
		\hline
		{Channel taps deviation ($\sigma_h^2$)} & $0$ or $0.1$ \\
		\hline
		{Impulse noise levels ($N$)} & $1$ or $2$ \\
		\hline
		{Impulsive index ($A$)} & $0.8$ \\
		\hline
		{Backgr.-to-impulse noise ratio ($\Gamma$)} & $0.01$ or $0.1$ \\
		\hline
		{Total noise power ($\sigma^2$)} & varied\\
		\hline
		{Impulse correlation ($r$)} & 0.98\\
		\hline
		{Number of detector states ($Q$)} & $N|\mathcal{X}|^{L}$  or $|\mathcal{X}|^{L}$ \\
		\hline
		\multirow{3}{*}{NN layers}
		& {$1\times100$ (Sigmoid)} \\
		& {$100\times50$ (ReLU)} \\
		& {$50\times Q$ (Softmax)} \\
		\hline
		{NN optimizer} & {Adam (learning rate $=0.01$)} \\
		\hline
		{GD iterations (NN)} & {20000} \\
		\hline
		{EM iterations (HMM)} & {1500} \\
		\hline
	\end{tabular}
	\label{tab:sim_param}
\end{table}

Table~\ref{tab:sim_param} lists the simulation parameters. We used a rate $1/2$ binary $(171,133)$ convolutional code with memory depth of 7 bits~\cite{NASA_code} which is standard for digital broadcasting systems. As in~\cite{Nir_BCJRNet} and~\cite{Ndo_MiddletonMarkov}, we considered the BPSK modulation format with transmitted symbol alphabet $\mathcal{X}=\{-1,1\} $. The transmitted sequence consisted of $5\cdot10^5$ symbols and $500$ different transmissions were simulated when evaluating the BER of the communication system at a specific SNR/SINR. The impulsive noise parameters were adopted from~\cite{Ndo_MiddletonMarkov}. Nevertheless, depending on the investigated communication scenario, channel parameters were varied accordingly. The neural network for the BCJR-NN detector followed the three layer architecture from~\cite{Nir_BCJRNet} with sigmoid, rectified linear unit~(ReLU) and final softmax activation. Stochastic GD optimization of the NN was performed over 20000 iterations using the Adam optimizer with learning rate of 0.01. The number of EM algorithm iterations for optimizing the HMM parameters was set to 1500.

\subsection{Intersymbol Interference Channel with AWGN}
\begin{figure}
	\centering
	\includegraphics[width=\columnwidth, keepaspectratio=true]{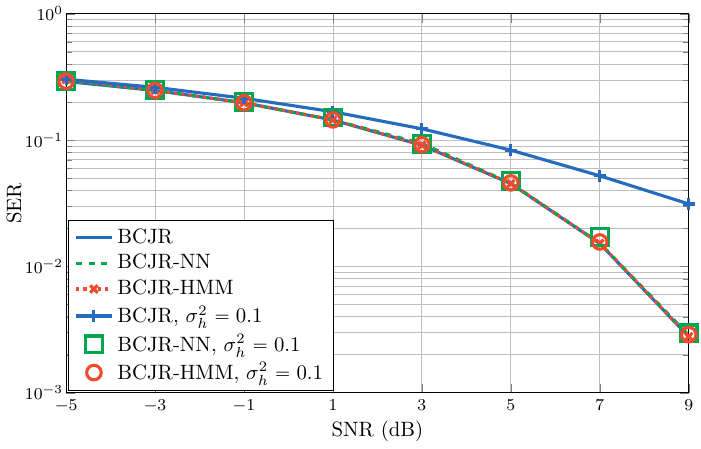}
	\caption{\label{fig:BPSK_ISI_AWGN_L2} SER versus SNR for uncoded BPSK transmission over a time-invariant ISI ($L=2$ and $\sigma^2_h=0$) channel with AWGN for systems with channel model-based or data-driven BCJR detection.}
\end{figure}

\begin{figure}
	\centering
	\includegraphics[width=\columnwidth, keepaspectratio=true]{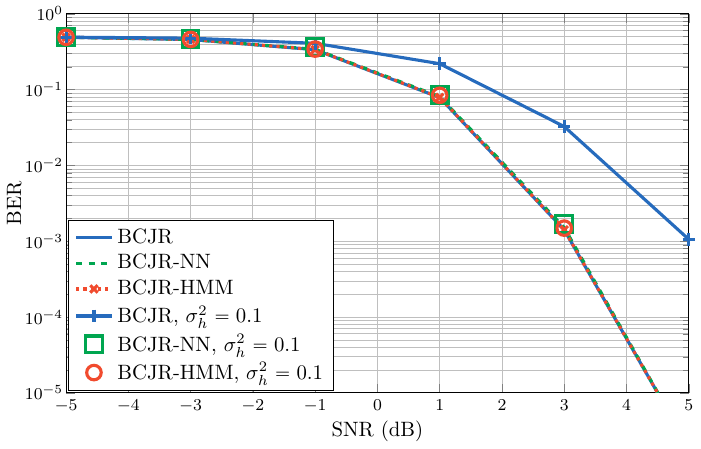}
	\caption{\label{fig:BPSK_ISI_AWGN_L2_coded}BER versus SNR for coded BPSK transmission over a time-invariant ISI ($L=2$ and $\sigma^2_h=0$) channel with AWGN for systems with channel model-based or data-driven BCJR detection.}
\end{figure}
We start by evaluating the symbol error rate (SER)  performance of the BCJR detectors for uncoded transmission over the ISI channel (symbol memory of $L=2$) and AWGN ($N=1$). In particular, as also done in~\cite{Nir_BCJRNet}, we considered the error rate performance for transmission over the time-invariant ISI and obtained the channel likelihoods in two different ways: 1) using the same time-invariant ISI model for both the full CSI-based BCJR and the optimization of the data-driven detectors and 2) using the rapidly time-varying model from~\eqref{eq:smean_} with $\sigma^2_h=0.1$ for this (referred to as \emph{CSI uncertainty} in~\cite{Nir_BCJRNet}). Note that for such a channel the BCJR state transitions, given by~\eqref{eq:ISI_tr}, are trivial and can be readily provided as prior knowledge for the channel model-based and NN detectors. Figure~\ref{fig:BPSK_ISI_AWGN_L2} shows the obtained SERs as a function of SNR for the different detection methods. As first reported in~\cite{Nir_BCJRNet}, we see that the BCJR-NN achieves SER performance close the optimal full CSI-based BCJR detector regardless of whether it was trained on the time-invariant or the time-varying channels. Our results show that the BCJR-HMM exhibits similar robustness to channel variations. In contrast, the SER of the channel model-based BCJR where likelihoods are computed with $\sigma^2_h=0.1$ is significantly degraded at higher SNRs. 

We extend the investigation of symbol detection over such channels for the case of coded transmission with the $(171,133)$ convolutional code. Figure~\ref{fig:BPSK_ISI_AWGN_L2_coded} shows the BER versus SNR curves for systems employing the investigated symbol detectors. For this setting we can also observe that the BCJR-NN and the BCJR-HMM detection schemes achieve BERs close to the algorithm with perfect CSI regardless of whether they were optimized via the time-invariant or the time-varying ISI model with $\sigma^2_h=0.1$. Furthermore, at 1\,dB SNR the data-driven solutions outperform by a factor of 2 the BCJR detection with inaccurate CSI --- an improvement which becomes more pronounced as the SNR increases.

\subsection{Impact of Detector States on the System Performance}
\begin{figure}
	\centering
	\includegraphics[width=\columnwidth, keepaspectratio=true]{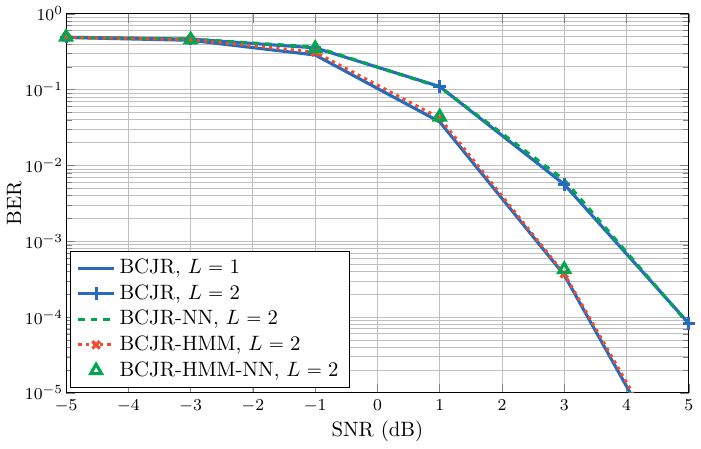}
	\caption{\label{fig:BPSK_AWGN_Rec_L2}BER versus SNR for coded BPSK transmission over an AWGN channel for BCJR detection based on different number of states.}
\end{figure}
\begin{table*}[!t]
	\caption{Likelihood function parameters and state transition matrix for BPSK over an AWGN channel ($10$\,dB SNR) learned in three different independent trials (left, center and right) using a 4-state HMM. Steady-state distributions are also listed. Conversely,~\eqref{eq:ISI_tr} needs to be used for the other $L=2$ detectors.\label{tab:State_trans_ISI_AWGN_oe}}
	\centering
	\begin{adjustbox}{max width=0.32\textwidth}
		\begin{tabular}{|c|c|c|c|c|}
			\hline
			\cellcolor{black!30!white}$\hat{\mu}$& \cellcolor{black!0!white}$-1$ & \cellcolor{black!0!white}$-1$ & \cellcolor{black!0!white}$1$ & \cellcolor{black!0!white}$1$ \\
			\hline
			\cellcolor{black!30!white}$\hat{\sigma}^2$& \cellcolor{black!0!white}$0.1$ & \cellcolor{black!0!white}$0.1$ & \cellcolor{black!0!white}$0.1$ & \cellcolor{black!0!white}$0.1$ \\
			\hline
			\cellcolor{black!15!white}\diagbox[width=\dimexpr \columnwidth/8+2\tabcolsep\relax, height=0.5cm]{$\mathbf{s}_{t-1}$}{$\mathbf{s}_t$} & \cellcolor{black!15!white}$\mathbf{s}^{1}$ & \cellcolor{black!15!white}$\mathbf{s}^{2}$ & \cellcolor{black!15!white}$\mathbf{s}^{3}$ & \cellcolor{black!15!white}$\mathbf{s}^{4}$ \\
			\hline
			\cellcolor{black!15!white}$\mathbf{s}^{1}$& 0.33& 0 & 0 & 0.67 \\
			\hline
			\cellcolor{black!15!white}$\mathbf{s}^{2}$& 0.25 & 0.49 & 0.13 & 0.13\\
			\hline
			\cellcolor{black!15!white}$\mathbf{s}^{3}$ & 0.3 & 0.2 & 0.32 & 0.18\\
			\hline
			\cellcolor{black!15!white}$\mathbf{s}^{4}$  & 0.29 & 0.21 & 0.16 & 0.34 \\
			\hline
			\cellcolor{black!15!white}$\pi$& \cellcolor{black!15!white}0.295 & \cellcolor{black!15!white}0.204 & \cellcolor{black!15!white}0.127 & \cellcolor{black!15!white}0.374\\
			\hline
		\end{tabular}
	\end{adjustbox}
	\hfill
	\begin{adjustbox}{max width=0.33\textwidth}
		\begin{tabular}{|c|c|c|c|c|}
			\hline
			\cellcolor{black!30!white}$\hat{\mu}$& \cellcolor{black!0!white}$-1$ & \cellcolor{black!0!white}$-1$ & \cellcolor{black!0!white}$1$ & \cellcolor{black!0!white}$1.7$ \\
			\hline
			\cellcolor{black!30!white}$\hat{\sigma}^2$& \cellcolor{black!0!white}$0.1$ & \cellcolor{black!0!white}$0.1$ & \cellcolor{black!0!white}$0.1$ & \cellcolor{black!0!white}$0.0001$ \\
			\hline
			\cellcolor{black!15!white}\diagbox[width=\dimexpr \columnwidth/8+2\tabcolsep\relax, height=0.5cm]{$\mathbf{s}_{t-1}$}{$\mathbf{s}_t$} & \cellcolor{black!15!white}$\mathbf{s}^{1}$ & \cellcolor{black!15!white}$\mathbf{s}^{2}$ & \cellcolor{black!15!white}$\mathbf{s}^{3}$ & \cellcolor{black!15!white}$\mathbf{s}^{4}$ \\
			\hline
			\cellcolor{black!15!white}$\mathbf{s}^{1}$& 0.06& 0.7 & 0.24 & 0\\
			\hline
			\cellcolor{black!15!white}$\mathbf{s}^{2}$& 0.02 & 0.47 & 0.51 & 0\\
			\hline
			\cellcolor{black!15!white}$\mathbf{s}^{3}$ & 0.03 & 0.47 & 0.5& 0\\
			\hline
			\cellcolor{black!15!white}$\mathbf{s}^{4}$  & 1 & 0 & 0 & 0 \\
			\hline
			\cellcolor{black!15!white}$\pi$& \cellcolor{black!15!white}0.026 & \cellcolor{black!15!white}0.476 & \cellcolor{black!15!white}0.498 & \cellcolor{black!15!white}0\\
			\hline
		\end{tabular}
	\end{adjustbox}
	\hfill
	\begin{adjustbox}{max width=0.32\textwidth}
		\begin{tabular}{|c|c|c|c|c|}
			\hline
			\cellcolor{black!30!white}$\hat{\mu}$& \cellcolor{black!0!white}$-1$ & \cellcolor{black!0!white}$-1$ & \cellcolor{black!0!white}$-1$ & \cellcolor{black!0!white}$1$ \\
			\hline
			\cellcolor{black!30!white}$\hat{\sigma}^2$& \cellcolor{black!0!white}$0.1$ & \cellcolor{black!0!white}$0.1$ & \cellcolor{black!0!white}$0.1$ & \cellcolor{black!0!white}$0.1$ \\
			\hline
			\cellcolor{black!15!white}\diagbox[width=\dimexpr \columnwidth/8+2\tabcolsep\relax, height=0.5cm]{$\mathbf{s}_{t-1}$}{$\mathbf{s}_t$} & \cellcolor{black!15!white}$\mathbf{s}^{1}$ & \cellcolor{black!15!white}$\mathbf{s}^{2}$ & \cellcolor{black!15!white}$\mathbf{s}^{3}$ & \cellcolor{black!15!white}$\mathbf{s}^{4}$ \\
			\hline
			\cellcolor{black!15!white}$\mathbf{s}^{1}$& 0.28& 0.72 & 0 & 0\\
			\hline
			\cellcolor{black!15!white}$\mathbf{s}^{2}$& 0.005 & 0.345 & 0.65 & 0\\
			\hline
			\cellcolor{black!15!white}$\mathbf{s}^{3}$ & 0 & 0.05 & 0.27& 0.68\\
			\hline
			\cellcolor{black!15!white}$\mathbf{s}^{4}$  & 0 & 0.13 & 0.37 & 0.5\\
			\hline
			\cellcolor{black!15!white}$\pi$& \cellcolor{black!15!white}0.001 & \cellcolor{black!15!white}0.129 & \cellcolor{black!15!white}0.369 & \cellcolor{black!15!white}0.501\\
			\hline
		\end{tabular}
	\end{adjustbox}
\end{table*}
Next, we considered a practically-relevant communication scenario where the number of detector states is over-estimated, for example to accommodate operation in a ``worst-case'' ISI scenario. In particular, we devise a simple setting of coded BPSK transmission over an AWGN channel where the BCJR symbol detectors are provisioned for ISI with $L=2$, i.e., they have $Q=|\mathcal{X}|^L=4$ states. Figure~\ref{fig:BPSK_AWGN_Rec_L2} compares the BER performance of the different detection schemes in such a scenario. As a reference baseline we use the channel model-based $L=1$ BCJR detector where states are the two BPSK symbols with transition probabilities of~$0.5$. Compared to the reference system, we observe a degraded  BER performance for the conventional BCJR as well as the BCJR-NN both operated with $L=2$. This is due to the fact that these detectors rely on the channel model-based computation of the state transition probabilities from~\eqref{eq:ISI_tr}, which is performed with the inaccurate ISI knowledge of $L=2$ and hence is not matched to the actual transmission channel. On the other hand, our results show that the BER performance of the BCJR-HMM is not significantly affected by the inaccurate number of detector states. Since the BCJR-HMM detection scheme uses the received symbol sequence to optimize both the parameters of the channel likelihood function as well as the state transition matrix, it manages to mitigate the effects of over-estimating the number of states by accordingly adjusting the state transitions. To emphasize this, Table~\ref{tab:State_trans_ISI_AWGN_oe} shows examples of the optimized state transition matrices and likelihood distribution parameters which we obtained in three independent optimization attempts. On the left-hand side we see that the learning algorithm separated the four available states into two groups of two with means of the channel likelihood distribution equivalent to the BPSK symbols (i.e., $\hat{\mu}=-1$ for $\mathbf{s}^1$ and $\mathbf{s}^2$, and $\hat{\mu}=1$ for $\mathbf{s}^3$, $\mathbf{s}^4$) and variances equal to the inverse of the linear SNR. Importantly, we observe that only certain transitions between states were allowed. To gain further insight on the optimized state transition matrix, the table also shows the corresponding steady-state distribution $\pi$ which allows us to see that the two groups have total probabilities of approximately $0.5$ which is in correspondence with the state transitions in the optimal $L=1$ detector. Similar grouping behavior can also be recognized in the middle and right-hand side examples. Note that for the former, the optimization converged to an unused state $\mathbf{s}^4$ with an outlier mean and very small variance with no incoming transitions allowed, while for the latter, three states were grouped together. 

To further examine the importance of the state transition matrix, we utilized the NN-based channel likelihood estimator with state transitions optimized via HMM. The BER performance curve when employing this BCJR-HMM-NN scheme is also shown in Fig.~\ref{fig:BPSK_AWGN_Rec_L2}. Interestingly, we observe that such a detection approach improves the performance of the BCJR-NN and allows operation close the the optimal full CSI-based BCJR due to the optimized transition matrix. Such a joint HMM-NN scheme can be particularly promising in applications with non-trivial channel likelihood functions where the powerful computational capabilities of the NN are particularly beneficial.

\subsection{Intersymbol Interference Channel with Bursty IN}

\begin{figure}
	\centering
	\includegraphics[width=\columnwidth, keepaspectratio=true]{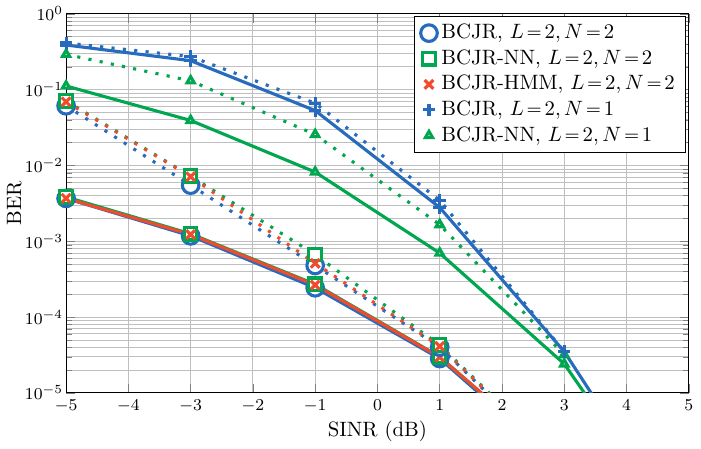}
	\caption{\label{fig:BPSK_ISI_IN_1}BER versus SINR for coded BPSK transmission over a time-invariant ISI ($L=2$ and $\sigma^2_h=0$) channel with bursty IN ($N=2$) for systems with channel model-based or data-driven BCJR detection. Solid lines correspond to $\Gamma=0.01$, while dotted lines to $\Gamma=0.1$.} 
\end{figure}

In this section we consider coded transmission over ISI~($L=2$) channel with bursty IN~($N=2$). The BER performance for the proposed systems with different detectors as a function of the SINR is shown in Fig.~\ref{fig:BPSK_ISI_IN_1}. To investigate different regimes of impulsive interference, the figure shows results for two different background-to-interference noise ratios $\Gamma=0.01$ and $\Gamma=0.1$. More specifically, we consider the cases where the number of states in the BCJR-HMM and the BCJR-NN detectors are matched to the number of channel states ($L=2,N=2$) and compare their performance with the channel model-based BCJR with perfect CSI ($L=2,N=2$) as well as conventional BCJR detection designed for the ISI channel with AWGN (the \emph{BCJR, $L=2,N=1$} system). We see that for the ISI with IN channel the BERs of the two data-driven receivers follow closely that of the optimal BCJR detection for both values of $\Gamma$. Conversely, there is a strong degradation in the system performance if detection is performed under the ISI with AWGN assumption, i.e., where only the total noise variance $\sigma^2$ is considered in computing the likelihoods and the state transitions are governed solely by the ISI process. The receivers for digital broadcasting systems are often based on the AWGN assumption and crucially the proposed machine learning solutions provide an opportunity to tailor their performance in a data-driven manner when impulsive interference is present. While the BCJR-HMM detector manages to fully adapt to the channel characteristics without prior knowledge of the IN, the BCJR-NN requires labeling of the IN states for training as well as providing the state transition matrix. These requirements could be impractical since the sources of impulsive interference are typically external to the communication system. To further investigate the scenario of employing neural networks for the ISI channel with IN, we consider the BCJR-NN with $L=2,N=1$ where states are designed assuming only the ISI effect, while the training data for the NN-based likelihood estimator is generated by the ISI channel with IN. To simulate this scenario we labeled only the ISI states during the supervised learning. The performance of such a system, which does not rely on prior knowledge of the IN, is also shown in Fig.~\ref{fig:BPSK_ISI_IN_1}. 
\begin{figure}
	\centering
	\includegraphics[width=\columnwidth, keepaspectratio=true]{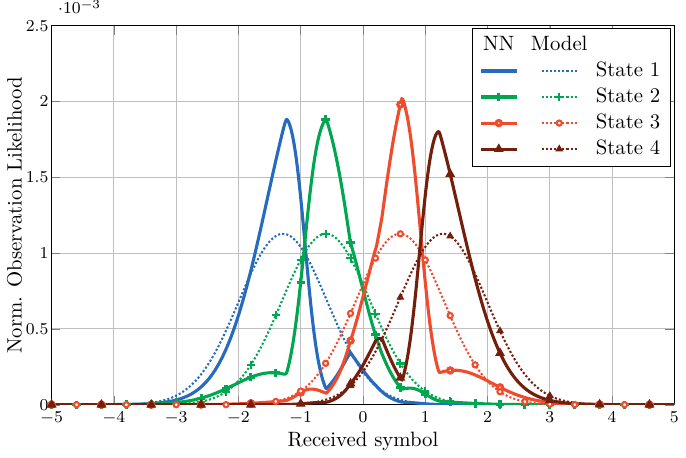}
	\caption{\label{fig:NN_model_obs}Normalized channel (observation) likelihoods for the time-invariant ISI ($L=2$ and $\sigma^2_h=0$) channel with IN~($N=2$) computed by the BCJR-NN and the conventional BCJR detectors with $L=2,N=1$.}
\end{figure}
Our results indicate that, while inferior to the detectors which accurately handle the IN states, the BCJR-NN ($L=2,N=1$) outperforms the conventional BCJR with $L=2,N=1$. When the number of states is reduced, the Gaussian assumption for the channel likelihoods is not necessarily accurate, hence the NN-based computation which learns the distribution directly from data provides a better estimate. To exemplify this, Fig.~\ref{fig:NN_model_obs} illustrates the (normalized) channel likelihoods obtained by the reduced-state NN and conventional AWGN detection methods at $3$\,dB SINR for $\Gamma=0.01$. Indeed, we see that the NN-learned distribution deviates from a Gaussian for each of the $Q=|\mathcal{X}|^L=4$ states. Another interesting observation from the BER results in Fig.~\ref{fig:BPSK_ISI_IN_1} is that the gains achieved by the reduced-state BCJR-NN over its conventional counterpart are smaller for $\Gamma=0.1$ compared to the case of $\Gamma=0.01$. In the extreme case of $\Gamma=1$ the Gaussian assumption for the likelihoods for these 4-state detectors becomes accurate since the background and interference noises have identical variances. This results in identical BER performance which we verified in our simulations. 

\begin{figure}
	\centering
	\includegraphics[width=\columnwidth, keepaspectratio=true]{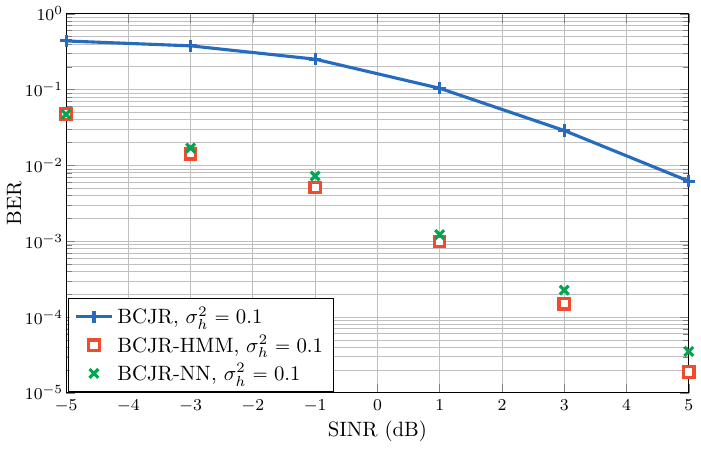}
	\caption{\label{fig:BPSK_ISI_IN_2}BER versus SINR for coded BPSK transmission over a rapidly time-varying ISI ($L=2$ and $\sigma^2_h=0.1$) channel with IN ($N=2$).}
\end{figure}

Next, we consider the impact of rapidly time-varying ISI behavior on the detectors' performance under bursty IN. Note that such a scenario for the ISI with AWGN channel and NN-based BCJR detection was considered as a practical use case in~\cite{Chen_1} (denoted by the authors as Case~5). Similarly here, to simulate such a system in presence of IN we set the deviation of the ISI taps to $\sigma^2_h=0.1$ for both the optimization and transmission channels. Figure~\ref{fig:BPSK_ISI_IN_2} shows the BER performance as a function of SINR for the BCJR-HMM and BCJR-NN detectors and compares them to the conventional BCJR where channel likelihoods were also computed with $\sigma^2_h=0.1$. Note that all detectors were provided with the accurate $Q=N|\mathcal{X}|^L=8$ number of states for this investigation. Compared to the BER performances over the time-invariant ISI model with IN, shown previously, our results indicate a considerable degradation in the examined systems performances due to the rapid ISI variations. Nevertheless, we notice that the data-driven detectors can provide significantly lower BERs at all examined SINRs compared to the conventional BCJR approach which cannot adapt to the changes in the ISI channel coefficients. Such an increased robustness to channel variations exhibited by the data-driven approaches is of great practical importance and thus presents a major direction in our future work.

\section{Conclusions}\label{sec:concl}

We considered two fundamentally different machine learning approaches --- based on the supervised learning of neural networks as well as the unsupervised optimization of hidden Markov models, for the implementation of data-driven BCJR symbol detection in communication scenarios characterized by non-trivial state transitions. To model these systems we focused on transmission over ISI channels with bursty IN as well as symbol detection with over-estimated trellis states in AWGN. Our results showed that obtaining the state transition matrix in a data-driven manner via the HMM framework is important for avoiding significant performance penalties. On the other hand, in settings where the channel likelihoods follow a non-Gaussian distribution, for e.g., when the IN states are not accounted for in the detector trellis, the NN-based estimator shows superior performance compared to the conventional detector operating on the AWGN assumption. Moreover, our work demonstrates the advantageous performance of a hybrid HMM-NN-based approach to learning the state transitions and the channel likelihoods for BCJR. Importantly, we extended our investigation of the data-driven detectors to rapidly time-varying ISI in the presence of IN and showed that these methods are significantly more robust to channel variations compared to the conventional detector.

\end{document}